\documentclass[aps, amssymb, amsmath, superscriptaddress, 
twocolumn] {revtex4-2}

\usepackage{chngcntr}

\usepackage{graphicx}
\usepackage{color}
\usepackage{amsmath}
\usepackage{enumitem}
\usepackage{amssymb}
\usepackage{hyperref}
\usepackage{cancel}
\usepackage{ulem}%%%%%%%
\usepackage{multirow}
\newcommand{\be}{\begin{equation}}
\newcommand{\ee}{\end{equation}}

\newcommand{\bea}{\begin{eqnarray}}
\newcommand{\eea}{\end{eqnarray}}

\newcommand{\p}{\partial}

\newcommand{\ra}{\right\rangle}
\newcommand{\lb}{\left[}
\newcommand{\rb}{\right]}
\newcommand{\lp}{\left(}
\newcommand{\rp}{\right)}

\newcommand{\Tr}{{\rm \, Tr\,}}

\renewcommand{\vec}[1]{{\boldsymbol #1}}

\makeatother

\begin{document}
	\title{Chiral Stoner magnetism in Dirac bands}
	%\author{Zhiyu Dong, Leonid Levitov}
	%\affiliation{Department of Physics, Massachusetts Institute of Technology, Cambridge, MA 02139}
%	\author{Zhiyu Dong}
%	\affiliation{Department of Physics, Massachusetts Institute of Technology, Cambridge, MA 02139}	
%	\affiliation{Department of Physics and Institute for Quantum Information and Matter, California Institute of Technology, Pasadena, California 91125}
	\author{Zhiyu Dong, Leonid Levitov}
	\affiliation{Department of Physics, Massachusetts Institute of Technology, Cambridge, MA 02139}
	
	\begin{abstract}
		%We argue that 
		Stoner magnetism in bands endowed with Berry curvature is shown to be profoundly influenced by the coupling between spin chirality density $\vec s\cdot(\p_x\vec s\times\p_y\vec s)$ and Berry's orbital magnetization. The key effect is that carriers moving in the presence of a spin texture see it as a source of a geometric magnetic field coupled to the carrier's orbital motion through a spin-dependent Aharonov-Bohm effect. This emergent spin-orbit interaction effect was recently predicted to enable chiral magnons propagating along system boundaries. Here we show that it also favors chiral spin textures such as skyrmions---the topologically protected objects with particle-like properties, stabilized in the ground state. The threshold for Stoner instability is found to soften, rendering chiral spin-ordered phases accessible under realistic conditions. We present a detailed analysis of the chiral effect for Bernal bilayer graphene and discuss the unique properties of skyrmion textures in graphene multilayers. %, yet the results are applicable to generic Stoner magnets. 
	\end{abstract}
	%\date{\today}
	
	\maketitle
	
%	The question of how Berry curvature impacts the many-body physics gained much attention with the advent of flat-band systems that host topological bands 
%	and strong interactions. 
The arrival of electron systems hosting topological flat bands and strong interactions
\cite{Lopes_dos_Santos,Mele,SuarezMorell,Bistritzer,Guinea} raises fascinating questions about the impact of Berry curvature on many-body physics.
%	The advent of electron systems which host topological flat bands and strong interactions 
%				\cite{Lopes_dos_Santos,Mele,SuarezMorell,Bistritzer,Guinea}
%	poses fascinating questions about the  impact of the Berry curvature on the many-body physics.  
%	poses many interesting questions for theory. Among the questions, perhaps
Recent work has focused on graphene multilayers such as moir\'e graphene \cite{Andrei2020,Cao2018Correlated,Cao2018SC,Cao2020,Zondiner2020,Wong2020, Saito2021}, where flat bands emerge when the moir\'e twist angle is tuned to a magic value \cite{Bistritzer}. Additionally, research has explored field-biased non-moir\'e bilayers and trilayers\cite{Zhou2022,Seiler2021,de la Barrera2022,Zhou2021isospin,Zhou2021SC}, systems in which bands are flattened by a transverse electric field \cite{McCann2013}. The small kinetic energy of carriers in flattened bands and the strong electron interactions characteristic of graphene create a platform in which a variety of ordered many-body phases can be realized and explored.
%	Recent work focused on 
%	graphene multilayers such as moir\'e graphene\cite{Andrei2020,Cao2018Correlated,Cao2018SC,Cao2020,Zondiner2020,Wong2020,Saito2021} in which flat bands appear when the moir\'e twist angle is tuned to a magic value \cite{Bistritzer}, as well as field-biased non-moir\'e bilayers and trilayers \cite{Zhou2022,Seiler2021,de la Barrera2022,Zhou2021isospin,Zhou2021SC}, systems in which bands are flattened by a transverse electric field \cite{McCann2013}. Small kinetic energy of carriers in a flattened band and strong electron interactions characteristic of graphene, create a platform in which a variety of ordered many-body
%phases can be realized and explored. 	%New exotic orders appear when 
	%or when bands are flattened by a transverse electric field \cite{McCann2013}. 
%	The diversity of the observed orders, which include insulating and superconducting phases coexisting with cascades of magnetic phases polarized in isospin (spin and valley) suggests seeking new interactions and many-body orders in these systems.
	The wide variety of observed orders, including cascades of magnetic phases polarized in isospin (spin and valley) alongside the insulating and superconducting phases, prompts seeking new interactions and previously unknown many-body orders in these systems. 
	%enabled by Berry curvature in flat bands. 
	%polarized phases
	%motivates 

	With this motivation, here we consider itinerant magnetic metals with spontaneously spin-polarized carriers %that occurs via Stoner instability mechanism 
	in bands equipped with Berry curvature. %Our analysis demonstrates that %carriers in 
	We find that such systems possess 
	%	the orbital %self-rotation 
	%degrees of freedom of carriers originating from Berry curvature are coupled to spin degrees of freedom by 
	a geometric %interaction between 
	coupling of the orbital and spin degrees of freedom  
	%and spin degrees of freedom 
	that favors nonzero spin chirality $\vec s\cdot (\p_1\vec s\times \p_2\vec s)$, where $\vec s(x)$ is the spin density and $\p_{1,2}$ are spatial derivatives. 
	%% Namely, we show that chiral interactions that can drive chiral spin orders are naturally present in interacting electronic systems %with bands that are equipped with due to the Berry curvature of carrier bands. 
	%Unlike the mechanisms studied before, our chiral interactions do not require any extra broken symmetries, microscopic SOI or frustration
	Such geometric coupling, enabled by Berry curvature and electron exchange interactions, is allowed to exist by general symmetry arguments. Below we develop a microscopic theory of this effect, which a particular focus on %predicts such effect for 
	%predicted by symmetry will be argued to exist %derived below for %are predicted to naturally occur in 
	itinerant graphene flat-band systems in which carriers are polarized in spin and/or valley (the $1/2$-metal and $1/4$-metal phases) as well as partially isospin polarized (PIP) phases. 
	The underlying physics %involves 
	can be understood as electromagnetic coupling $-\vec M\vec B$ between orbital magnetization $\vec M$ due to the band Berry curvature and the geometric magnetic field $\vec B$ originating from spin chirality. From a more general point of view, 
	the effect that can be thought of as an emergent geometric spin-orbit interaction (SOI) driven by electron exchange, 
	with the coupling strength taking a universal value mandated by geometric and topological constraints. %In essence, in these systems, 
%	the general question of how 

\begin{figure}
		\centering
		\includegraphics[width=1.0\linewidth]{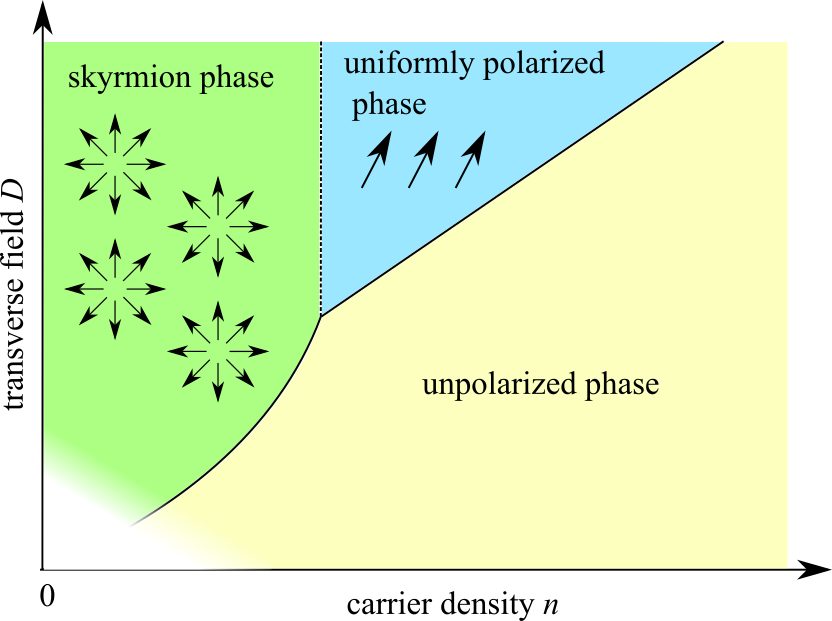}
		\caption{The mean-field phase diagram for chiral magnetic order 
			in the quadratic Dirac model, Eqs.\eqref{eq:H_total} and \eqref{eq:H_BBG}, away from charge neutrality at $D=0$. Transition between uniformly spin-polarized and unpolarized phases is first-order, occurring on a straight line 
			in the $n$-$D$ plane given by Eq.\eqref{eq:Stoner transition}. Transition between skyrmion phase and uniformly polarized phase is second-order, with 
			skyrmion density vanishing at the phase boundary given by 
			Eq.\eqref{eq:threshold2}. Phase boundaries are found 
			by focusing on spin polarization in one valley and ignoring valley ordering (see text).  
		}
		\label{fig:1}
		%\vspace{-3mm}
	\end{figure}

The geometric SOI established in this work has several notable implications. One is the existence of new collective excitations---chiral magnons propagating along the edges of the system \cite{dong2023collective}. These chiral spin waves, induced by the geometric SOI, are expected to emerge robustly in all itinerant spin-polarized phases in bands with Berry curvature. Another implication of the geometric SOI is the emergence of new magnetic phases in which spins form textures such as spin-density waves or skyrmions. Below, after demonstrating the effect microscopically, we analyze the stability of a uniformly polarized phase. We find that in the presence of geometric SOI, it becomes unstable towards formation of skyrmion textures at sufficiently low carrier densities (see Fig.\ref{fig:1}).

%	Point out that these orders occur in the presence of a p-space Berry curvature and pose the general question of how Berry curvature changes magnetism. Orbital magnetism, and anomalous Hall effect. What's the impact on magnetism? 

	The interplay between orbital effects and spin degrees of freedom is a topic that has been widely studied in magnetism, where microscopic spin-orbit interactions have been %shown to lead 
	used to 
	%LL produce 
	stabilize different kinds of helical and chiral orders in magnetic systems. 
	%Improve flow...} 
	Of special interest are the chiral %itinerant 
	magnetic phases, 
	wherein spins wrap around the Bloch sphere. 
	%LL, spanning a solid angle. 
	Chiral spin textures have been explored 
	in various magnetic systems\cite{Nagaosa2013, Fert2017}. Famously, the Dzialoshinskii-Moriya (DM) coupling favors spin textures\cite{Bogdanov2001} 
	such as helical spin density waves \cite{Binz2006, Nakanishi1980,Kataoka1981} and  
	skyrmions---the seminal topologically-protected particle-like spin configurations\cite{Polyakov1987}. 
	One can therefore ask whether it is possible to achieve a chiral spin order in graphene-based systems. At first sight, 
	this may seem problematic as staple SOI interactions that stabilize chiral magnetic orders are absent or extremely weak in graphene.
	Indeed, in noncentrosymmetric magnets---the bulk chiral magnetic metals\cite{Roessler2006,Muhlbauer2009,Neubauer2009,Yu2010,Seki2012,Schultz2012} and magnetic layers\cite{Heinze2011,Jiang2017}---skyrmions are stabilized by the DM coupling  
	governed by the microscopic spin-orbit interactions %(SOI)
	\cite{Bogdanov2001,Fert2017}.
	However, the microscopic SOI in graphene is usually negligible compared to other energy scales\cite{Neto2009}. Likewise, the mechanisms that utilize frustration\cite{Okubo2012,Leonov2015,Hayami2017} are not found in 
	graphene systems. Yet, as will be demonstrated below, the 
	%\sout{emergent} 
	geometric SOI originating from the interplay of a band curvature and spin polarization favors chiral spin orders without invoking any extra broken symmetries, microscopic SOI or frustrations. 
	
%	In this paper, we address this challenge by demonstrating a new mechanism that favors chiral spin orders. Namely, 
%	we show that 
%	chiral interactions that can drive chiral spin orders are naturally present in interacting electronic systems with bands that are equipped with Berry curvature. 
%	Unlike the mechanisms studied before, our chiral interactions do not require any extra broken symmetries, microscopic SOI or frustrations. 
	
%LL
\section{symmetry of the chiral effect} % symmetry}

The quantity that will be central to our discussion is the chirality density, or topological density, of a spin texture $\vec s(\vec x)$ defined as
a scalar triple product
\be\label{eq:chirality_triple_product}
b(\vec x)=\frac12\epsilon_{\mu\nu}\vec s(\vec x)\cdot(\partial_\mu \vec s(\vec x)\times \partial_\nu \vec s(\vec x))
,
\ee
where $\epsilon_{\mu\nu}$ a $2\times2$ antisymmetric tensor. This quantity, which characterizes twisting in the texture, has several remarkable properties. First, despite being a function of angles which cannot be reduced to a function of the modulus $|\vec s|$, it is invariant under SU(2) spin rotations. Second, it is odd under time reversal and spin-reversal operations, $t\to-t$ and $\vec s\to-\vec s$. The SU(2) symmetry property suggests that chirality could emerge from spin exchange interactions that depend on the angles between interacting spins and obey SU(2) symmetry. However, being odd under time and spin reversal symmetries excludes this quantity from basic models of magnetism such as the Heisenberg model and its extensions. 
%This behavior is quite different from what is usually encountered in a Heisenberg model of magnetism. 
Yet, as we will see, this quantity is mandated by symmetry for magnetism in bands with Berry curvature. 

%To derive this emergent interaction and explore its implications, we consider the problem of itinerant spin magnetism  
Specifically, despite its exotic symmetry properties, this quantity appears naturally in the problem of itinerant spin magnetism in Dirac bands such as those found in graphene multilayers, moir\'e \cite{Song2015, Bistritzer,Andrei2020} and non-moir\'e\cite{McCann2013,Castro2008}. In these systems carriers reside in valleys $K$ and $K'$ that are mapped to one another under time reversal symmetry. at the same time, the band dispersion and interactions in each valley individually are not constrained by time and spin reversal symmetries. As a result, the symmetry constraints for the quantities such chirality are relaxed. Allowed by the symmetries of the microscopic Hamiltonian, these quantities must be included in the theoretical description of the magnetic order. Furthermore, the analysis described below, which links these quantities to geometric gauge fields, predicts a universal value of the coupling constant that is large enough to significantly alter the physics of the ordered phase. 
%The chiral interaction strength is found to be large enough to change considerably the physics of the ordered phase. As we will see, the interaction expressed as chirality density, Eq.\eqref{eq:chirality_triple_product}, can soften the threshold for Stoner instability and stabilize chiral spin textures in the ordered phase. Fig.\ref{fig:1} illustrates this for a quadratic Dirac band model of a field-biased bilayer graphene (BBG) analyzed below.
%The strength of the chiral interaction is significant enough to considerably impact the physics of the ordered phase. 
As we will see, the interaction expressed as chirality density in Eq. \eqref{eq:chirality_triple_product} can lower the threshold for Stoner instability and stabilize chiral spin textures in the ordered phase. Figure \ref{fig:1} illustrates this for a quadratic Dirac band model of field-biased bilayer graphene (BBG) analyzed below.
	%\addQ{[If there is really such ``spin inversion", then it should be odd for each individual valley, killing the chiral interaction. However, I think the issue is that there is just no such spin-inversion symmetry because the operator inverting spin has to be anti-unitary, which essentially contains a time-reversal operator that has to simultaneously act on orbital part of wavefunction. Do you agree? ]} 
	%\addQ{\sout{is of a chiral character that %, results in a chiral interaction which leads to Stoner instability towards chiral ordered states. } \sout{triggers a chiral spin texture in itinerant magnets.} [changed it because a)``instability" sounds like we are studying phase transition, b)I would like not to call chiral spin texture an ``order" as it does not break symmetry.]}  

	The chiral interaction responsible for this behavior
	%\sout{that drives this \addQ{\sout{ordering} texture} }
	arises from the coupling between orbital magnetization due to $k$-space Berry curvature of Dirac carriers and %\sout{. This interaction occurs in the presence of a} 
	position-dependent spin polarization, 
	taking the form:
	\be\label{eq:deltaF_MB}
	\delta F=\int d^2x \sum_{i=K,K'} -(M_{i,+}-M_{i,-}) B_i(\vec x)
	,
	\ee
	where $M_{i,\pm}$ is the orbital magnetization of the majority-spin and minority-spin carriers in valleys $K$ and $K'$ (see Eq.\eqref{eq:Es}). This interaction can be viewed as an extension of the basic electromagnetic 
	coupling of a magnetic moment and external field, $E=-\vec M\cdot\vec B$, where
	the `magnetic field' $B_i(\vec x)$ is defined as topological density of spin texture in each valley multiplied by the flux quantum $\phi_0$:
	\be\label{eq:B=chirality}
	B_i(\vec x) =\frac{\phi_0}{4\pi }  \vec s_i\cdot(\partial_1 \vec s_i\times \partial_2 \vec s_i)
	, \quad \phi_0= hc/e, 
	\ee
	with $\vec s_i(x)$ the unit-vector field representing spin polarization of carriers in valleys $i=K,K'$. The quantity $B_i(\vec x)$ represents a geometric `magnetic field' associated with the spin-dependent (geometric or chiral) 
	Aharonov-Bohm effect. 
	This effect originates from the fundamental properties of %key aspect of the chiral interaction, Eq.\eqref{eq:deltaF_MB}, is that in the 
	spin textures $\vec s_i(\vec x)$ in which the position-dependent quantization axis along which carrier spins are polarized is, in general, allowed to twist in space. Spins of carriers moving through a texture and coupled to it by exchange undergo adiabatic spin rotation, which leads to geometric phase and magnetic field $B_i(\vec x)$ associated with the Aharonov-Bohm effect due to this phase. The magnetic field originates from carrier movement through the texture, is coupled to orbital degrees of freedom of carriers, leading to the the $-\vec M\vec B$ coupling to orbital magnetization of carriers appearing in Eq.\eqref{eq:deltaF_MB}. 
	%Eq.\eqref{eq:deltaF_MB}, derived below, can be viewed as an extension of the basic electromagnetic coupling of a magnetic moment and external field, $E=-\vec M\cdot\vec B$.
	
	Since $M_i$ describes orbital magnetization whereas $B_i$ equals %a property 
	(up to a factor) to the chirality of a spin texture, the interaction in Eq.\eqref{eq:deltaF_MB} can be viewed as an emergent spin-orbit interaction which originates from spin exchange and does not rely on a microscopic spin-orbital coupling. To understand the symmetry properties of this interaction, it is instructive to compare it to 
%superficially resembles 
the atomic spin-orbital interaction (SOI) $E\sim \vec L\cdot \vec s$. The SOI interaction locks spin polarization $\vec s$ to the orbital angular momentum vector $\vec L$. As a result, it is not invariant under SU(2) spin rotations. To the contrary, the chiral interaction originates from SU(2)-invariant spin exchange interactions, which endows it with unique symmetry properties distinct from those of SOI. 
	Indeed, the interaction in Eq.\eqref{eq:deltaF_MB} is governed by the interplay of the direct-space geometric phase which gives rise to $B_i(\vec x)$ and momentum-space Berry curvature responsible for $M_{i,\pm}$.  %exchange interaction, 
	%not relying on spin-orbital coupling. It therefore has 
The quantities $B_i$ are manifestly invariant under the $SU(2)$ spin rotations performed separately in each valley, whereas the orbital quantities $M_{i,\pm}$ remain invariant under such spin rotations. As a result, the chiral interaction, while describing a coupling between the spin and orbital degrees of freedom, is invariant under spin rotations. This is a high symmetry which is not known for the spin-orbit interactions originating from a microscopic spin-orbital coupling.

Furthermore, the interaction in Eq.\eqref{eq:deltaF_MB} also respects discrete symmetries of BBG, including the time-reversal and mirror symmetries. Indeed, the time-reversal symmetry maps $\vec s_K$ to $-\vec s_{K'}$, therefore it maps $B_K$ to $-B_{K'}$. Simultaneously, the orbital magnetization $M_K$ is mapped to $-M_{K'}$. Consequently, Eq.\eqref{eq:deltaF_MB} is invariant under time-reversal symmetry.  Likewise, the mirror symmetry maps $M_K$ to $-M_{K'}$, whereas $B_K$ is mapped to $-B_{K'}$. Indeed, taking for example the mirror lying in $yz$ plane, the gradient components $(\partial_1,\partial_2)$ are mapped to $(-\partial_1,\partial_2)$, leaving the chirality density in Eq.\eqref{eq:chirality_triple_product} unchanged. Consequently, Eq.\eqref{eq:deltaF_MB}  respects mirror symmetry. Same conclusion applies to the other two mirrors as they are equivalent to the $yz$ mirror up to a $C_3$ rotation. 

We note that in this reasoning, due to the absence of microscopic SOC, the spin and orbital degrees of freedom are decoupled and we do not need to account for the mirror symmetry action on spin degrees of freedom. However, it is straightforward to show that the conclusion would remain the same if we accounted for mirror symmetry action on spin. Indeed, the $yz$ mirror maps $\vec s_K = (s_K^x,s_K^y,s_K^z)$ to  $(s_{K'}^x,-s_{K'}^y,-s_{K'}^z)$, which leaves the chiral density invariant. 
Further, we note parenthetically that the $C_2$ symmetry that interchanges the top and bottom graphene layers 
%\addQ{[CORRECT?]} 
is broken in the systems of interest due to the presence of a transverse electric field. This is a crucial condition for the system to host the chiral coupling because otherwise (in the presence of the $C_2$ symmetry) the orbital magnetization would not be allowed by symmetry and must vanish.

%\addLL{%Our first goal is to 
%Here we put our analysis in a general symmetry framework, separating the symmetry properties of the Hamiltonian and of the order parameter in the symmetry-broken phase.
%}% Discuss separately H,  n, and alpha. }

%In this section, we check that 
This discussion demonstrates that, while the `emergent SOI interaction' given in Eq. \eqref{eq:deltaF_MB} may appear exotic, it follows the form allowed by symmetry. This observation carries two important implications, one almost trivial and the other considerably less trivial. First, while the coupling in Eq. \eqref{eq:deltaF_MB} obeys the continuous spin rotation symmetry and discrete time reversal and spin inversion symmetries, %all or some of 
these symmetries may be broken in the system ground state. As illustrated by Fig.\ref{fig:1}, this is indeed what happens in the system of interest. Second, all interactions of the form allowed by symmetry are expected to be supported by a generic realistic Hamiltonian. Therefore, regardless of the details of the microscopic derivation presented below, all conclusions of our analysis can be inferred and justified based on symmetry grounds alone. This observation makes the results of the analysis presented below applicable broadly and valid outside the specific assumptions under which the analysis will be carried out. 

We note parenthetically that from the symmetry point of view, the properties of electron bands in rhombohedral multilayer graphene are the same as those in BBG. Therefore, the analysis carried out for BBG is applicable to a wider variety of multilayer graphene systems, such as rhombohedral-stacked trilayer and pentalayer graphene. 
%\addQ{Let's broaden this a bit. What about ABC trilayers and pentalayers?} 
	
It is also worth nonting that the chiral interaction that stabilizes skyrmions in our theory %essentially different from the one described in recent works, 
is distinct from the effects considered in the literature. In particular, skyrmions in isospin-polarized moir\'e graphene flat bands have been invoked to predict exotic superconductivity \cite{Bomerich2020, Khalaf2021}. The mechanism that stabilizes skyrmions in these papers is an isospin extension of quantum Hall ferromagnet physics, in which the skyrmion emerge in Landau levels spin-split by exchange interactions \cite{Sondhi1993, MacDonald1996, Fertig1997, Nagaosa2013}. Skyrmions of this type have been predicted \cite{Yang2006,Nomura2006} and recently observed \cite{Zhou2020,Liu2022} in graphene at high magnetic fields. 
	
%\section{The symmetry requirements of the chiral interaction}\label{sec:symmetry}

%{\bf In this section, we do a symmetry analysis of the chiral interaction to show that the chiral interaction robustly arises. }
%find that chiral interaction occurs in the system that spontaneously breaks the same symmetry as that of the applied magnetic field. }

%We are interested in a chiral interaction that appears even without an applied magnetic field. 
	
\section{Itinerant magnetism in a band with Berry curvature} %, a microscopic model} %a Dirac band}
	\label{sec:model}
	Here, we discuss Stoner magnetism in a graphene bilayer Dirac band. This analysis will set the stage for deriving microscopically the interaction between spin-chirality and orbital magnetization enabled by Berry curvature, Eq.\eqref{eq:deltaF_MB}. %, by a microscopic analysis 
	Our starting point is a fully SU(2)-invariant Hamiltonian involving a single-particle Hamiltonian and electron-electron interactions, but not involving any microscopic SOI, 
	\be\label{eq:H_total}
	{\cal H}= \mathcal H_0 +\mathcal H_{\rm{int}}.
	%\sum_p c_p^\dagger H(p) c_p
	%-\frac12 \sum_{\vec x,\vec x'} U(\vec x-\vec x') : s_\alpha(x) s_\alpha(x'):
	%,
	\ee	
We take the single-particle part $\mathcal{H}_0$ to be a quadratic Dirac Hamiltonian of a Bernal-stacked graphene bilayer: %motivated by graphene following Hamiltonian:}
	\be
	\mathcal{H}_0 = \sum_{\eta,p} c_{\eta,p}^\dagger H_\eta(p) c_{\eta, p}
	\ee	
	where  $H_\eta(p)$ is a $2\times 2$ Hamiltonian
	\be\label{eq:H_BBG}
	H_\eta(p)=\lp
	\begin{array}{cc}
		D & \frac{(p_1- i\eta p_2)^2}{2m}\\
		\frac{(p_1+ i\eta p_2)^2}{2m} & -D
	\end{array}
	\rp
	\ee
	Here, $\eta=\pm 1$ for the valleys $K$ and $K'$; %respectively. Accordingly, 
	the quantities %$\psi(x)$, $\psi^\dagger(x)$, 
	$c_{\eta, p}$, $c^\dagger_{\eta, p}$ are spinors with the  $A$ and $B$ sublattice components and the ordinary spin-$1/2$ components.
	%The quadratic Dirac band is described by a general $2\times 2$ Dirac Hamiltonian in the sublattice $A$, $B$ basis. 
	%Here we consider the quadratic Dirac problem 
%The $2\times 2$ 
The Hamiltonian $H_\eta(p)$ possesses particle-hole symmetry, 
with the effects of particle-hole asymmetry and trigonal warping ignored for simplicity. 
Incorporating these terms later or 
generalizing to other Dirac band types 
would be straightforward. Estimates for realistic parameter values 
are provided in Sec.\ref{sec:energetics of spin textures} (see discussion beneath Eq.\eqref{eq:threshold2}).

Next, we consider the electron-electron interaction. For simplicity, we focus on the intra-valley spin exchange and ignore the exchange interaction between electrons in valleys $K$ and %lectrons in valley 
	$K'$. The inter-valley  processes are expected to be weak as they % is negligible since it requires 
	involve a large momentum transfer, and are therefore subleading to the intra-valley processes. Indeed, %has a negligible strength since 
	the electron-electron interaction predominantly arises from Coulomb interaction,  which decreases as $1/p$ %the inverse of 
	as a function of the momentum transfer $p$. 
Restricting the exchange interaction to electrons in the same valley, we introduce %can model the interaction between them using 
a spin-exchange coupling of the form
	\be\label{eq:H_int}
	{\cal H}_{\rm{int}}= 
	%\sum_\eta -\frac12 \sum_{\vec x,\vec x'} U(\vec x-\vec x') : s_{\eta\alpha}(x) s_{\eta\alpha}(x'):
	 -\frac12 \sum_{\eta,\vec k} U(\vec k) : s_{\eta\alpha}(\vec k) s_{\eta\alpha}(-\vec k):
	.
	\ee 
	Here the exchange interaction written in terms of spin density %$s_\alpha(x)=\psi^\dagger(x)\sigma_\alpha \psi(x)$ 
	%$s_{\eta,\alpha}(x) = c^\dagger_{\eta,x}\sigma_\alpha c_{\eta,x} $
	$s_{\eta \alpha}(k) = \sum_p c^\dagger_{\eta,p+k}\sigma_\alpha c_{\eta,p} $
	with Pauli matrices $\sigma_\alpha$ representing ordinary spin-$1/2$ variables, $\alpha=1,2,3$. 
%	\addQ{This effective spin-exchange coupling microscopically originates from a bare exchange coupling $V(q)$ that arises from the Coulomb interaction, and the renormalization by itinerant fermions. The bare exchange coupling can be model using a contact interaction $V(q) = V_0$. The renormalization effects by itinerant fermions can be accounted for through 
		%the fermion's spin susceptibility $\chi(q)$. Namely, the strength of this effective spin-exchange interaction can be expressed as
%		the following expression\cite{Coleman2015} 
%	\be 
%	U(q) = \frac{V_0}{1-V_0\chi(q)},
%	\ee 
%	where $\chi(q)$ represents the fermion's spin susceptibility. At small $q$, the spin susceptibility $\chi$ can be expressed as $\chi(q) = \chi_0 \lp 1 - \xi_0^2 q^2\rp$ where $\xi_0$ is a characteristic length scale determined by the band structure of itinerant carriers. Consequently, the effective spin-exchange coupling $U(q)$ takes the following form at small $q$,
%	\be \label{eq:U(q)}
%	U(q) = U_0 \lp 1 - \xi^2 q^2\rp ,\quad U_0 = \frac{V_0}{1-V_0\chi_0}, \quad \xi = \xi_0\sqrt{U_0 \chi_0}.
%	\ee 
%}
%	The density-density interaction $e^2/\kappa |r-r'|$, which generates the exchange interaction on a microscale, is suppressed for conciseness. Its role will be discussed below. 
%Here 
Microscopically, the spin-exchange coupling originates from the Coulomb interaction. However, directly starting from microscopic %Coulomb 
interactions will significantly complicate the analysis. Here, to illustrate the physics of interest within a simplest formulation, we replace the microscopic exchange interaction with a toy-model dependence %the toy-model form 
\be
U(\vec k)=U_0 e^{-k^2\xi^2}
,
\ee
where $\xi$ is a correlation lengthscale.

Equivalently, the spin exchange Hamiltonian 
%of exchange energy 
given in Eq.\eqref{eq:H_int} can be written in coordinate representation as
\be	
{\cal H}_{\rm{int}}= \sum_\eta -\frac12 \sum_{\vec x,\vec x'} U(\vec x-\vec x') : s_{\eta\alpha}(\vec x) s_{\eta\alpha}(\vec x'):
, 
\ee
%where $U(q)$ is the Fourier transform of a 
with a nonlocal spin-spin interaction
%will use 
%with a toy-model form of exchange coupling, 
\be
	U(\vec x-\vec x')=4\pi U_0 \xi^{-2} e^{-(\vec x-\vec x')^2/4\xi^2}
\ee
	normalized %so that 
	to $\int d^2 x U(\vec x)=U_0$. As we will see, tuning the interaction radius $\xi$ provides a convenient knob for studying the spin-polarized phase, analyzing fluctuations and charting the phase diagram of skyrmion textures.
	%where the last term is 

 In our model, given by Eqs. \eqref{eq:H_BBG} and \eqref{eq:H_int}, the electrons in two valleys are decoupled. Therefore, in the analysis below we can consider the $K$ valley alone. %($\eta=1$). 
 It should be noted that in reality the electrons in valleys $K$ and $K'$ also interact through a direct density-density interaction (the Hartree energy). However, we do not need to include this interaction  in the model because this term only gives a charging energy that determines the total electron density. As such, it does not affect the spin polarization in each valley, which is quantity of interest.

	\section{mean field theory for spin textures}
	\label{sec:meanfield}
	To describe spin textures, we
	perform a mean-field analysis  
	in which the field describing ensemble-averaged spin polarization 
	is allowed to vary in space. Since the exchange interactions are predominantly intravalley it will be sufficient to carry out the analysis for an individual valley and consider the role of valley degrees of freedom later. The Hubbard-Stratonovich (HS) transformation is carried out using an ordering field $\vec h(\vec x)$ with both the modulus and orientation being position-dependent, 
	\begin{align}
	& \exp\lp \int dt \sum_k\frac{U(\vec k)}{2}\vec s_{\vec k}\cdot \vec s_{-\vec k}
	\rp
	\\ \nonumber
	&=\int D[\vec h] \exp\lp \int dt\sum_{\vec k} \vec h_{\vec k}(t)\cdot \vec s_{-\vec k}-\frac{\vec h_{\vec k}(t)\vec h_{-\vec k}(t)}{2U(\vec k)}\rp 
	, \end{align}
	where $D[\vec h]=\prod_{\vec k,t}d\vec h_{\vec k}(t)$. Here we introduced Fourier harmonics of the HS field and spin density % and the interaction 
	$\vec h_k=\int d^2x \vec h(\vec x)e^{-i\vec k\vec x}$, $\vec s_k=\int d^2x \vec s(x)e^{-i\vec k\vec x}$. %, $U(k)=\int d^2 xU(\vec x) e^{-i\vec k\vec x}= U_0e^{-k^2\xi^2}$.
	Integrating out fermions and assuming a time-independent $\vec h(\vec x)$, we obtain the free energy with a nonlocal $h(\vec x)h(\vec x')$ interaction 
	\begin{align}
	\label{eq:F1}
	F=
	\Tr \log\lb i\omega-H(p)-h_\alpha(\vec x) \sigma_\alpha\rb
	+\sum_{\vec k} \frac{\vec h_{\vec k}\vec h_{-\vec k}}{2U(k)}
	,
	\end{align}
	where, for conciseness, the chemical potential $\mu$ is incorporated in $H$ and  $\Tr$ denotes $\sum_{\vec x}\int \frac{d\omega d^2p}{(2\pi)^3} \Tr_{2\times 2}$.  Eq.\eqref{eq:F1} is an exact result for the fermion partition function, with no approximations made, which is applicable for any position-dependent ordering field $\vec h(\vec x)$. 
	
	In this framework, % it is easy to compare 
	the behavior of the states with uniform polarization and those with a general space- and time-dependent $\vec h(\vec x,t)$ can be compared on equal footing. 
	The saddle point condition $\delta F=0$ yields a time-independent $h=|\vec h|$, which is 
	nothing but the Stoner mean field value
	\be\label{eq:Stoner}
	h=U(0)(n_+-n_-)/2,
	\ee
	where $n_+$ and $n_-$ are local densities of carriers with spins parallel and antiparallel to local spin quantization axis $\vec h(x)$. We will call these spin species the majority and the minority spins, respectively. 
	When the system is fully polarized, the mean field equals $h=U(0)n/2$.

	Next, we consider weakly inhomogeneous $\vec h(\vec x)$. The term $-h_\alpha(\vec x)\sigma_\alpha$ describes electron spins coupled to a spin texture with a position-dependent magnetization polarized along the unit vector $\vec s(\vec x)=\vec h(\vec x)/h$, where $|\vec h(\vec x)|=h$. We therefore write $\vec h(\vec x)=\vec h_0+\delta \vec h(\vec x)$, where $\delta \vec h(\vec x)\perp\vec h_0$, and approximate the dependence of the free energy on $\delta\vec h(\vec x)$ as a second-order functional derivative
\be
\delta F=F[\vec h(\vec x)]-F[h_0]=\sum_{\vec k}\frac12\frac{\p^2 F}{\p \vec h_{\vec k}\p \vec h_{-\vec k}}\delta \vec h_{\vec k}\delta \vec h_{-\vec k},
\ee 
where $F[h_0]$ is the %logarithmic term 
free energy evaluated for spatially uniform $\vec h(\vec x)$. 
Expanding the logarithmic term in the free energy [Eq.\eqref{eq:F1}] in $\delta \vec h(\vec x)$ to second order yields:
\begin{equation}\label{eq:second-order_deltah}
\delta F=%F[h_0]+
\sum_{\vec k} \frac{1}{2} \chi_{\pm}(\vec k)\delta \vec h_{\vec k}\cdot\delta \vec h_{-\vec k} +\frac{\delta\vec h_{\vec k}\cdot\delta\vec h_{-\vec k}}{2U(\vec k)},
\end{equation}
where $\chi_{\pm}(k)$ is the Lindhard function of spin-polarized Fermi sea. 
%Here $\delta F$ is a second-order functional derivative
%\be
%\delta F=F[\vec h(x)]-F[h_0]=\frac12\frac{\p^2 F}{\p\delta \vec h_{\vec k}\p\delta \vec h_{-\vec k}}\delta \vec h_{\vec k}\delta \vec h_{-\vec k},
%\ee 
%where $F[h_0]$ is the %logarithmic term 
%free energy evaluated for spatially uniform $\vec h(\vec x)$. 

This result can be used to evaluate spin stiffness. 
%\addQ{Through direct calculation of $\chi_{\pm}$ one finds $\chi_{\pm}(0)=1/U(0)$, which cancel with the $\vec k=0$ contribution from the last term. This agrees with the Goldstone's theorem which requires long-wavelength magnon to be gapless.}
In doing so, we expect that the $\vec k=0$ contributions of $\chi_{\pm}(k)$ and $1/U(k)$ cancel out, %with the $\vec k=0$ contribution from the last term, 
since only the spatially varying part of $\vec h(\vec x)$ contributes to the energy of a weakly inhomogeneous symmetry-broken state (as required by Goldstone's theorem).
%\addLL{
Accordingly, we consider the dependence $F$ vs. $\delta \vec h_{\vec k}$
%last term in the free energy [Eq.\eqref{eq:F1}] $F_{h}=\sum_k \frac{\delta\vec h_k\cdot\delta\vec h_{-k}}{2U(k)}$, 
assuming that the local spin-quantization axis is slowly varying, $\xi \partial_\mu S_\alpha \ll1$. 
This is the case when large spin stiffness makes %the exchange interaction radius $\xi$ exceeds the Fermi wavelength, making 
the short-wavelength fluctuations in $\vec h(\vec x)$ costly and, therefore, weak. 
%	\sout{Accordingly,  
%	we approximate $ U(k)^{-1}= U_{0}^{-1} (1+k^2\xi^2/2) $, which gives}
%\sout{\addQ{Using Eq.\eqref{eq:U(q)}, we obtain}}
Expanding in $k$ at second order, 
$U^{-1}(k)\approx U^{-1}(0) (1+k^2\xi^2) $, % and expanding 
$\chi_{\pm}(k)\approx \chi_{\pm}(0)+ak^2$, %+O(k^4)$,
gives the dependence of the free energy on $\delta \vec h_{\vec k}$:
	\be\label{eq:F_{h}}
	%F_{h}=\sum_k \frac{\vec h_{k}\cdot \vec h_{-k}}{2U(k)}\approx 
	\delta F=%F[h_0]+
	\sum_{\vec k} \frac12 \lp \chi_{\pm}(0)+ak^2 +\frac{1+k^2\xi^2}{U(0)}%+\frac1{2}\frac{\xi^2 \addQ{U(0)}} k^2 
	\rp \delta \vec h_{-\vec k}
\delta \vec h_{\vec k}	.
	\ee
	%Plugging this in Eq.\eqref{eq:second-order_deltah}, we expand $\chi_{\pm}(k)=\chi_{\pm}(0)+ak^2+O(k^4)$ and 
	Taking into account that $\chi_{\pm}(k)$ is evaluated for the spin-polarized state, we confirm that  $k=0$ contributions cancel out owing to the Stoner mean-field relation, Eq.\eqref{eq:Stoner}. This yields the gradient expansion
	\be\label{eq:stiffness}
	%F[\vec h(x)]=F[h_0]+
	\delta F=\sum_x \frac1{2}J (\p_\mu S_\alpha)^2
	,\quad
	J= \frac{\xi^2 h^2}{U(0)} + ah^2
	.
	\ee
	This result indicates that the spin stiffness parameter $J$ is dominated by $U(k)$ expansion to order $k^2$ when the correlation length $\xi$ is large. For a contact interaction the stiffness is dominated by Lindhard function exapansion. For a realistic short-range interaction, both contributions to stiffness are expected to be equally important.% }
	%\addQ{Do we want to discuss the sign of $a$ and the possibility of negative stiffness?}} %$J=\frac{\xi^2 h^2}{\addQ{U(0)}}$.
	
	%The value of $a$ sensitively depends on the band structure. For simplicity, here and below we focus on the large-$\xi$ regime $\xi^2 \gg a U(0)$ so that the spin stiffness $J$ is positive-definite. The positivity of $J$ is required because otherwise spin density wave pattern would set in, invalidating the approach we are currently using above, which is the perturbation theory around the uniform saddle point. Therefore, below we will estimate the stiffness by 
The value and sign of $a$ are sensitive to the details of the band structure and may potentially result in a negative spin stiffness $J$. To ensure that $J$ remains positive, we focus here on the large-$\xi$ regime where $\xi^2 \gg a U(0)$. In this case, the stiffness value is dominated by the first term in Eq.\eqref{eq:stiffness}: 
		\be 
		J\approx \frac{\xi^2 h^2}{U(0)} .
		\ee
	The positivity of $J$ is crucial because otherwise, a spin density wave pattern would emerge, rendering the approach relying on perturbation theory around the uniform saddle point invalid. %Therefore, below, we estimate the stiffness as follows:

%	 \addLL{OK? Is this the only contribution to spin stiffness?}
	
	\section{spin-dependent geometric magnetic field} % and chiral interaction }
	%What is the role of the 
	Next, we extend the mean-field framework to describe  spin textures. This can be done by considering the spatial dependence of the spin polarization field $\vec h(\vec x)$. In this section, we show that the spin texture generates a geometric spin-dependent  magnetic field $B(\vec x)$ given by Eq.\eqref{eq:B=chirality}, which couples to electrons as described in Eq.\eqref{eq:deltaF_MB}.
		
	We first give a heuristic argument for the origin of the field $B(\vec x)$. Microscopically, it arises from the adiabatic spin rotation effect for spins of electrons moving through a long-period spin texture to which they are coupled by exchange interaction. % is rotated in spin space. 
	The spin rotation describes evolution of an electron spin being locked to the local spin quantization axis and tracking it along the electron trajectory. In the adiabatic regime, the effect can be described by a spin-dependent geometric phase that depends on position-dependent spin polarization in the texture. This adiabatic framework is applicable %regime in which the geometric phase picture applies occurs 
	when the Stoner spin gap is large compared to $\hbar v/\ell$, where $\ell$ is the characteristic spatial lengthscale of the spin texture modulation and $v$ is the Fermi velocity. Berry curvature associated with the vector potential describing this geometric phase, defines a geometric magnetic field 
	\be\label{eq:b_geometric}
	\vec b(\vec x)=\nabla_{\vec x}\times\vec a(\vec x)
	.
	\ee 
Below, we identify the quantities $\vec a$ and $\vec b$ with %. To this end, we derive the 
	spin-dependent gauge fields obtained from the microscopic Hamiltonian and demonstrate that $\vec b$ equals to the chirality density given in Eq.\eqref{eq:chirality_triple_product}. From this, we derive the geometric coupling between $\vec b$ and electrons' orbital magnetization, which is the interaction given in Eq.\eqref{eq:deltaF_MB}.
	%we substantiate  the physical picture of a spin-dependent gauge field by a microscopic derivation. 
	
	%Below, we show by a microscopic derivation that the spin textures, described by the spatial dependence of spin polarization $h(x)$, generates a spin-dependent pseudo magnetic field. 

	%Next, to clarify the origin 

	Next, we formally introduce a gauge field describing a position-space Berry phase for electrons in the presence of a slowly varying spin texture. In that we follow the procedure developed some time ago in the literature on quantum antiferromagnets and high-temperature superconductivity \cite{Baskaran1988,Wiegmann1988,Schulz1990,Ioffe1991} and, more recently, in the literature on frustrated magnetic systems \cite{Kenya2000,Fujita2011,Hamamoto2015}. 
		%\addQ{For reader's convenience,} 
		Below we present a step-by-step derivation of the gauge field picture %\sout{results given in Eqs.\eqref{eq:a_mu} and \eqref{eq:chirality_density}}
		%spin-dependent gauge field and the chiral interaction in Eqs. \eqref{eq:deltaF_MB} and \eqref{eq:B=chirality}, %we derive step by step 
	starting with the microscopic Hamiltonian introduced in Sec. \ref{sec:model}. %Eq.\eqref{eq:B=chirality}. 
	In doing so, it will be shown explicitly that a spin texture gives rise to an effective gauge field whose flux density is associated with the spin chirality. In our analysis below, without loss of generality, we focus on spins in valley $K$ and suppress the valley label. A spin texture is described by a position-dependent ordering field $\vec h(r)$ introduced in Sec.\ref{sec:meanfield} through a Hubbard-Stratonovich mean field analysis, which we will write as
	\be\label{eq:texture}
	\vec h(\vec x)=h \vec s(\vec x)
	\ee
where	$\vec s(\vec x)$ is a unit-vector field, $|\vec s(\vec x)|=1$. In what follows, we will ignore fluctuations of the order parameter magnitude $h$ and focus on the fluctuations of $\vec s(\vec x)$ orientation in spin space.
	
	The first step is to perform an SU(2) spin rotation to bring all the local spin polarization to the same orientation and, in this way, generate a 
	Hamiltonian that features a geometric spin-dependent gauge field.
	We start with a one-electron Hamiltonian in valley $K$, writing it in position space:
	\be\label{eq:H}
	H(r) = \lp \begin{array}{cc}
		D & \frac{(p_1-  ip_2)^2}{2m}\\
		\frac{(p_1+  ip_2)^2}{2m} & -D
	\end{array} \rp 1_{\rm S}- h 1_{\rm L} \vec s(r)\cdot \vec \sigma
	\ee
	where $1_{\rm S}$ and $1_{\rm L}$ represent the identity matrices in the spin and sublattice subspaces, respectively. 
	In the first term, $p_{1,2}=-i\partial_{1,2}$ (here we set $\hbar=1$, restoring dimensional units later). The second term represents the effect of a position-dependent spin polarization arising after a Hubbard-Stratonovich transformation, 
	see Eq.\eqref{eq:F1}. 
	
	%--------------

	%This is done by 
	%Specifically, we carry out 
	The coordinate-dependent spin rotation operator $T(\vec x)$ that rotates all spins from the local polarization direction $\vec s(\vec x)$ to the $+\vec z$ direction is defined through 
		%Using this $T(x)$ operator, the spin-polarized state $|\vec s(x)\pm \rangle$ can be expressed as
		%perform a position-dependent $SU(2)$ similarity transformation $T(r)$ on the Hamiltonian, such that it rotates all spins to the $+z$ direction:
	%Spin rotation 
	%which is carried out at every point in position space, 
	\be\label{eq:z_axis}
	\left.|\vec z\pm \ra=T(\vec x)\left.|\vec s(\vec x)\pm \ra .
	\ee 
	%such that 
	%As a result, the operator $T(x)$ satisfies the following identity,
	When acting with this spin rotation on the Hamiltonian in Eq.\eqref{eq:H}, 
	%as expected, 
	the coordinate-dependent ordering field in the term $\vec h_\alpha(\vec x)\sigma_\alpha$ is transformed to a uniform field pointing in $+\vec z$ direction:
	\be
	h \sigma_3 =T(\vec x) \vec h_\alpha(\vec x)\sigma_\alpha  T^\dagger(\vec x).
	\ee 
	However, the simplicity comes at a price: the momentum operator in the Hamiltonian, Eq.\eqref{eq:H}, is transformed by $T$ to a long derivative with a $2\times2$ matrix gauge field. Namely, the Hamiltonian is transformed to
	%This transformation allows us to express the spatial-dependent Hamiltonian 
	%as a position-dependent $SU(2)$ similarity transformation $T(r)$ 
	%as a spin rotation from a uniform hamiltonian $H_z$ which describes a system where spins are uniformly polarized in $+z$ direction, }
	%the Hamiltonian Eq.\eqref{eq:H} can be expressed as
	\begin{align}
		\nonumber
		&H_z(\vec x) = T(\vec x) H(\vec x) T^\dagger(\vec x),
		\\
		&= \lp \begin{array}{cc}
			D & \frac{(\Pi_1-  i\Pi_2)^2}{2m}\\
			\frac{(\Pi_1+  i\Pi_2)^2}{2m} & -D
		\end{array} \rp 1_{\rm S}- h 1_{\rm L} \sigma_3\label{eq:H_z}
	\end{align}
	where 
	\bea
	&& \Pi_\mu  = - iT(\vec x)  \partial_\mu T^{\dagger}(\vec x) = p_\mu + A_\mu,  \nonumber\\
	&& A_{\mu} = - iT(\vec x)  \lb \partial_\mu , T^{\dagger}(\vec x) \rb, \quad \mu=1,2 \label{eq:A_mu}
	\eea
	Here $A_{1,2}$ are $2\times 2$ matrices representing an $SU(2)$ gauge field, and the square brackets represent commutators. The quantities $A_\mu$ can be expressed in terms of Pauli matrices:
	\be
	A_\mu = \sum_{i=1,2,3} a_{\mu, i}\sigma_i
	,
	\ee
	where the coefficients $a_{\mu, i}$ are the scalar quantities 
	%defined by Eq.\ref{eq:A_mu}.
	%\sout{that can be calculated through}
	\be\label{eq:a_mu}
a_{\mu,i}(\vec x)= \frac{1}{2}\Tr(\sigma_{i} A_\mu) = - \frac{i}{2}\Tr (\sigma_i T (\vec x)\partial_\mu T^{\dagger}(\vec x)).
\ee

This analysis, which is exact so far, simplifies in the adiabatic regime, where all spins 
	track the spin-up and spin-down states in rotated basis. In this case, the off-diagonal components  
	$a_{\mu,1}$ and $a_{\mu,2}$ describe coupling between spin states split by exchange. Such couplings only contribute at subleading order since they induce off-resonant transitions which are weak in the adiabatic limit. We can therefore  retain only the diagonal $\sigma_3$ 
	spin components, given by Eq.\eqref{eq:a_mu}. This gives %finding 
\be\label{eq: Pi=p+a}
	\Pi_\mu = p_\mu + %\sout{\addQ{\frac{e}{\hbar c}}}
	a_{\mu} \sigma_3,
	\ee 
	where from now on $a_{\mu}$ will be used as a shorthand for $a_{\mu,3}$. Plugging this back to Eq.\eqref{eq:H_z}, and absorbing $\mu$ in $H_z(p)$, we have the following form of free energy
		\bea\label{eq:F2}
		F=%\sum_{x,p,\omega} %\int d^2x \sum_{\omega,p} 
		\Tr \log\lb i\omega-H_z(\vec p + \vec a\sigma_3) -h \sigma_3\rb	+F_{h}
		\eea
		where $\Tr$ denotes $\int d^2x \sum_{\omega,p} \Tr_{2\times 2}$ 
		and $F_{h}$ is the spin stiffness energy given in Eq.\eqref{eq:F_{h}}.  
		
	%\sout{These results indicate that the spin-up and spin-down electrons \addQ{in the rotated basis}, which describe the majority and minority spin in the original basis, see  $U(1)$ gauge fields of opposite signs} 
	
The results indicate that the spin-up and spin-down electrons in the rotated basis, describing the majority and minority spin in the original basis, experience $U(1)$ gauge fields of opposite signs.
	%\addQ{\sout{ (here spin-up and spin-down refers to states in a rotated basis)}}. 
	After some algebra, which follows closely that in Ref.\cite{Fujita2011}, one finds
	\be
	a_{\mu} = -\frac{1}{2}(1-\cos \theta) \partial_\mu \phi 
	\label{eq:A}
	\ee
	where $\theta$ and $\phi$ are the spherical polar and azimuthal angles measured with respect to the $z$ axis introduced in Eq.\eqref{eq:z_axis}. 
The geometric magnetic field $\vec b=\nabla\times \vec a$ is then given by 
	\be\label{eq:b=chirality}
	b(\vec x)= \epsilon_{\mu\nu} \partial_{\nu} a_{\mu}=\nabla_x \times\vec a
	=%\frac{1}{2}
	%\addQ{\frac{\phi_0}{4\pi}}
	\frac12 \epsilon_{\mu\nu} \vec s\cdot(\partial_\mu \vec s\times \partial_\nu \vec s)
	,
	\ee 
%	\addQ{where $\phi_0 = \frac{2\pi \hbar c}{e}$.}
	which is nothing but the chirality density given in Eq.\eqref{eq:chirality_triple_product}. The result in Eq.\eqref{eq:A} indicates that the geometric phase picked up by an electron moving in the magnetic field $b(\vec x)$ is equal to $1/2$ of the solid angle swept by the spin quantization axis.

So far, we have utilized the `geometric' units where $\hbar=1$ and the units for $\vec p$ and $\vec a$ are inverse length. Consequently, the geometric field $\vec b$ is expressed in units of inverse length squared. It is interesting to examine how these quantities and their relationships alter when physical units are restored, setting the units for $\vec p_\mu$ and $\vec a_\mu$ to $\hbar$ over length.

Importantly, %We first note that 
the relationship between $\Pi_\mu$ and $p_\mu$ given in Eq.\eqref{eq: Pi=p+a} remains unchanged upon reintroducing physical $\hbar$ units. Furthermore, comparing this relation to the canonical electromagnetic coupling,
\begin{equation}\label{eq:Pi=p+a_EM}
\boldsymbol{\Pi} = \vec p - \frac{e}{\hbar c}\vec A,
\end{equation}
we observe that, while the coupling between the geometric field $\vec a$ and the electrons' orbital degrees of freedom mirrors the canonical electromagnetic coupling in form, the coupling strength is quite different
%. However, notably, the coupling strength differs significantly 
from the canonical coupling in Eq.\eqref{eq:Pi=p+a_EM}. The coupling to geometric gauge field leads to energy scales that are 
independent of $e$ and $c$. Instead, the energy scales are
governed by the lengthscales describing spatial periodicity of skyrmion textures (see Eq.\eqref{eq:Beff} and accompanying discussion). 
% . Specifically, the geometric coupling strength in Eq.\eqref{eq: Pi=p+a} is approximately $\hbar c/e^2\sim 137$ times stronger than the electromagnetic coupling strength in Eq.\eqref{eq:Pi=p+a_EM}.

\section{%chiral interaction from 
gradient expansion in $\vec a(\vec x)$}

The discussion above implies that the geometric magnetic field $\vec b=\nabla\times \vec a$ is perceived by electrons as a pseudo magnetic field coupled to orbital degrees of freedom with an effective strength given by:
\begin{equation}\label{eq:B_eff}
\vec B(\vec x)=\frac{\phi_0}{2\pi} \vec b(\vec x),
\end{equation}
where $\phi_0 = \frac{2\pi \hbar c}{e}$ is the flux quantum. This relationship arises from recognizing that Eq.\eqref{eq:b=chirality} indicates that a skyrmion texture with a unit topological charge,
\begin{equation}
N=\int d^2x \frac{1}{4\pi}\epsilon_{\mu\nu} \vec s\cdot(\partial_\mu \vec s\times \partial_\nu \vec s),
\end{equation}
translates into a unit flux quantum of an effective magnetic field perceived by an electron.

%, making this coupling strength approximately $\hbar c/e^2\sim 137$ times greater than the `naively expected' atomic-scale coupling strength.

%\addQ{The large strength of this coupling implies that skyrmion textures are capable of inducing  strong orbital current effects. One such effect is the anomalous Hall response. The anomalous Hall effect generated by spin texture is expected to be equivalent to the Hall effect generated by a huge magnetic field. This is similar to the observation in Nd$_2$Mo$_2$O$_7$\cite{Taguchi2001} where a non-coplanar spin texture is found to generate a gigantic anomalous Hall effect.}

This observation has direct implications for the coupling strength of the chiral interaction given 
%LL governing the chiral interaction introduced above 
in Eqs.\,\eqref{eq:deltaF_MB} and \eqref{eq:B=chirality}. 
%LL The significant 
The appreciable strength of this coupling implies that skyrmion textures are capable of inducing strong orbital current effects. One such effect is the anomalous Hall response. The estimates provided above indicate that the anomalous Hall effect generated by spin texture is equivalent to the Hall effect induced by a large magnetic field. This behavior resembles the behavior observed in Nd$_2$Mo$_2$O$_7$ \cite{Taguchi2001}, where a non-coplanar spin texture led to a significant anomalous Hall effect.

	We are interested in the instability of a spatially-uniform magnetic order 
	towards a twisted state with a nonzero gauge field $\vec a$. 
	We therefore consider power-series expansion of the electronic energy in Eq.\eqref{eq:F2} in small $\vec a$. 
	We neglect the longitudinal fluctuations of $\vec h$, which are gapped, focusing on the soft angular fluctuations, $\delta \vec h(\vec x)\perp \vec h$.
	For a slowly varying unit-vector field $\vec s(\vec x)=\vec h/|\vec h|$, the dependence on
	$\vec a$ in 
	the first term of Eq.\eqref{eq:F2}, hereafter referred to as $F_1$, is of the form given by the Peierls substitution $\vec p\to \vec p+\vec a\sigma_3$. 
	The expansion in  powers of $\vec a$, because of the gauge covariance of the free energy $F_1$, must involve gauge-invariant quantities expressed as gradients of $\vec a$ such as $\vec b=\nabla\times\vec a$. At first order in $\vec a$, the dependence on $\vec a$ can be linked to orbital magnetization by following a standard argument  from electromagnetic theory. Namely, for both the majority and minority spin projections, one can write the coupling between orbital currents and geometric vector potential as 
	%Once this is established for both minority and majority spins, one can immediately express the coupling as 
	$\delta H=-\frac{1}{c}\vec j\cdot \vec A$, and then rewrite it in terms of the geometric magnetic field given in Eq.\eqref{eq:B_eff} 
	as $-\vec M \cdot \vec B$ using  $\nabla\times \vec M = \frac{1}{c}\vec j$ and $\nabla \times \vec A=\vec B$.
	%\addQ{[do we need to fix dimensional factors, or keep it the way it is?]}. 
	%an expansion in powers of $\vec a$. 
	In a similar manner, we can write terms second order in $\vec a^2$ as $\frac12\chi B^2$, where $\chi$ is the orbital magnetic susceptibility. Putting everything together we arrive at
	\bea\label{eq:Es}
	&&F_1 = \sum_{\pm}E_\pm  -\Delta M B +\frac12\chi B^2, \quad \Delta M = M_+-M_-,\nonumber
	\\
	&&E_\pm=\sum_k(\epsilon^\pm_k-\mu) f(\epsilon^\pm_k),\quad \chi=\chi_+ + \chi_-. 
	\eea
	where the quantities
	$M_\pm$ and $\chi_\pm$ are the orbital magnetizations and the Landau diamagnetic susceptibility of the majority and minority spins,  
	in the presence of the pseudo magnetic field
	$B(\vec x) = \frac{\hbar c}{e}\nabla\times \vec a$ 
	 (see Eq.\eqref{eq:B_eff}). 
 %\addQ{\sout{Using the relation in Eq.\eqref{eq:chirality_density}, it is straightforward to show that $B$ defined in this way is identical to $B$ given in Eq.\eqref{eq:B=chirality}.}} 
	The $E_\pm$ contributions are the energies of spin-majority and spin-minority fermions in the bands with an exchange spin splitting,
	$\epsilon^s_k=\epsilon_k\mp h$, evaluated at $\vec a=0$, whereas the second and third terms represent the dependence on the pseudomagnetic field $\nabla\times \vec a$ at second order in $\vec a$. 
	
	The contributions $\mp M_\pm B$ 
	describe orbital magnetization of spin-majority and spin-minority carriers, arising due to Berry curvature, coupled to the pseudomagnetic field. 
	Crucially, both the conduction and valence bands contribute to $M$. 
	Therefore, perhaps counterintuitively, both up-spin and down-spin contributions to $M$ matter even if the conduction band is fully polarized. The values $M_\pm$ depend on the band filling and will be discussed below. The sign $\mp$ accounts for the fact that the Berry phase for the carriers with opposite spins, moving in a slowly varying texture $\vec h(\vec x)$, has opposite signs, described by the $\sigma_3$ factor in Eq.\eqref{eq:F2}. 
	In this form, Eq.\eqref{eq:F2} describes the limit of a weak, non-quantizing pseudomagnetic field 
	$B$, which is sufficient for the purpose of analyzing the transition from zero to nonzero 
	$B$.
	
%	\addQ{
%	The general question of how to express $M$ in terms of Hamiltonian $H$ is covered in the literature, see  Ref.?. Perhaps, what Referee A actually wants us to show is a derivation of the $H=-M B$  term from microscopics. Therefore, we need to show in a microscopic model that the spin texture couples to electron through the standard Peierls substitution form $p+A$ . Once this is established for both minority and majority spins, one can immediately express the coupling as $H=-j\cdot A$, and then rewrite it as $-M B$ using  $\nabla\times M = j$ and $\nabla \times A=B$. Similarly, the term $B$ follows directly from the Peierls substitution $p+A$ . To this end, in the resubmitted version, we present a microscopic derivation of the pseudo gauge field term $p+A$ in the appendix. This result follows closely previous literature (Ref.) but we still derive it in Supplement for reader's convenience.
%}

\section{Stoner instability in the absence of chiral interaction} %stability of spin-polarized phase} % in the presence of chiral interaction}
	Putting everything together, and for now ignoring the chiral interaction, we can write the system energy in the absence of pseudo magnetic field $B$ as
	\be\label{eq:F3} 
	F= \int d^2 x  \lb E_{+}+E_- +\frac{h^2}{2U(0)} 
	+\frac{J}{2}(\partial_\mu \vec s)^2\rb 
	\ee
	Using this expression, we can seek the ground state by comparing the energies of the ordered and disorder states. To account for the effect of a long-range $1/r$ density-density interaction without incorporating it explicitly in the mean-field analysis, we consider different states at the same total carrier density $n$. This approach is valid due to the large charging energy values $E_c = \frac12 V_0 n^2$ which typically exceed other energy scales in the system. When $E_c$ is included in the analysis, the dependence of the total energy on $n$ is dominated by the following terms:
	\be
	\frac{V_0 n^2}{2} -\mu n = \frac{V_0}{2}\lp n-\frac{\mu}{V_0}\rp^2-\frac{\mu^2}{2V_0}
	\ee
	These terms pin the density to $n=\frac{\mu}{V_0}$ regardless of the order type. Therefore, comparing energies of different states at the same $\mu$ in the presence of $E_c$ is equivalent to comparing their energies at the same $n$.
	
	To analyze the ordering described by Eq.\eqref{eq:F3} we proceed in two steps: First analyze the Stoner instability 
	while temporarily ignoring $B$ 
	. Next, we consider the dependence on $B$
	and the transition from a uniform magnetic order to a twisting order. 
	
	In the absence of $B$, Eq.\eqref{eq:F3} describes the standard Stoner instability---a transition from a disordered state to a uniformly polarized state. 
	Since the density of states in 
	the quadratic Dirac band monotonically decreases as a function of energy, the ground state configuration is either fully spin polarized or spin unpolarized, depending on the band parameters and interaction strength. 
	[For a more general band dispersion, partial spin polarization can also occur.]
	The energy density of a fully polarized phase with $n_+=n$ and $n_-=0$ is given by 
	\be
	F_{\rm{fp}} = E_{\rm tot}(n) - \frac{U_{0}n^2}{2}
	\ee
	where $n$ is a given total carrier density. We have used Eq.\eqref{eq:Stoner}. %and $E_\uparrow = E_{\rm tot}(n)-hn$
	Here 
	\be
	E_{\rm tot}(n)=\int_0^{\sqrt{4\pi n}} \frac{d^2k}{4\pi^2} \epsilon_k
	\ee 
	represents the total kinetic energy of electrons of density $n$ in one spin one valley in the absence of interaction.
	Similarly, the energy of unpolarized state where  $n_{+}=n_{-}=n/2$ is given by
	\be
	F_{\rm{unp}} = 2E_{\rm tot}(n/2)
	\ee
	Here, we have used $h=0$ in unpolarized phase.
	For our quadratic Dirac band, $E_{\rm tot}$ takes the following form:
	\be
	E_{\rm tot}(n) = \frac{mD^2}{4\pi} \lp \log(x+\sqrt{1+x^2})+x\sqrt{1+x^2}\rp , 
	\ee
	where $x=\frac{2\pi n}{mD}$ is a dimensionless density parameter.
	The regime of interest is that of strong exchange interaction, which corresponds to low values $n\ll 2mD$. In this case, we can use power-series expansion $E_{\rm tot}(n) = \frac{mD^2}{2\pi} ( x+\frac{5}{12}x^3 +
	...)$. 
	This allows a direct comparison of the energies of polarized and unpolarized states. 
	Simple algebra then predicts the fully polarized state to win when 
	\be
	\frac{n}{D} < \frac{2m^2 U_{0} }{5\pi^2}
	.\label{eq:Stoner transition}
	\ee
	Therefore, the phase boundary is a straight line on the $D$-$n$ phase diagram (see Fig.\ref{fig:1}). This phase transition is first-ordered since the full polarization occurs abruptly. 
	
	\section{chiral interaction and energetics of spin textures}\label{sec:energetics of spin textures}
	Next, we consider %the role of a 
	spin textures $\vec s(\vec x)$ and derive the condition for skyrmion proliferation in the presence of chiral interaction. From Eqs.\eqref{eq:Es},\eqref{eq:F3}, we see that system energy depends on $\vec s$ as
	$
	E_{\vec s}=\int d^2x \lb \frac{J}2  (\p_\mu S_\alpha)^2
	\mp \Delta M B
	+\frac{\chi}2 B
	^2\rb.
	$
	Therefore, the spin texture enters the energetics in two ways:  
	through pseudo magnetic field $B$ which is proportional to
	the spin chirality density Eq.\eqref{eq:b=chirality},
	and also through the spin stiffness energy $\frac12 J(\p_\mu \vec s)^2$. However, the latter contribution has a lower bound associated with spin chirality 
	\[
	\frac{1}{2}\int d^2x (\partial_\mu \vec s)^2 \geq \frac{1}{2}\int d^2x|\epsilon_{\mu\nu} \vec s\cdot(\partial_\mu \vec s\times \partial_\nu \vec s)|
	.
	\] 
	This relation follows from the well-known identity\cite{Polyakov1987}:
	\bea
	\int d^2x \lb (\partial_\mu \vec s)^2 \mp \epsilon_{\mu\nu} \vec s\cdot(\partial_\mu \vec s\times \partial_\nu \vec s)\rb
	\label{eq:identity_a}\\
	= \frac{1}{2}\int d^2x(\partial_\mu \vec s \pm \epsilon_{\mu\nu}\vec s \times\partial_\nu \vec s)^2 \geq 0,
	\label{eq:identity_b}
	\eea 
	Expressing 
	the stiffness energy through $|B|$ gives
	\be
	E_{\vec s}[B] = \int d^2 x \lb -\Delta M B + \frac{2J e}{\hbar c} |B| +\frac{\chi}{2} B^2\rb, \quad \label{eq:Fc}
	\ee
	Obviously, the quantity $E_{\vec s}[B]$ is only well-defined when spin polarization occurs. Therefore, below we focus on the effect of $E_{\vec s}[B]$ on the fully spin-polarized state. 
	
	It is now straightforward to derive the condition for nonzero chirality to be favored. The free energy in Eq.\eqref{eq:Fc} gives the threshold for nucleating chiral spin textures in the ground state:
	\be\label{eq:threshold}
	\Delta M \geq 2Je/\hbar c
	. 
	\ee
	As a reminder, $\Delta M =M_{+}-M_-$, 
	$M_{\pm}=M(\mu\pm h)$ in one particular valley. Below, without loss of generality, we focus on $K$ valley. 
	
	% THIS GIVES ONLY PART OF THE TOTAL MAGNETIZATION?
The net orbital magnetization of all electrons in one valley is governed by the Dirac band and its Berry curvature. This quantity, evaluated in our particle-hole-symmetric Dirac model, takes a simple form (see Ref.\cite{Xiao2007} and Appendix \ref{Appendix}):
	\be\label{eq:M}
	M_K(\mu) 
	= \begin{cases}
		\frac{eD}{2\pi \hbar c},\quad \mu>D\\
		\frac{e\mu}{2\pi \hbar c},\quad -D<\mu<D\\
		-\frac{eD}{2\pi \hbar c},\quad \mu<-D\\
	\end{cases}
	,
	\ee
	taking opposite values in valleys $K$ and $K'$.
This result is derived following the approach in Ref. \cite{Xiao2007}, where a similar result was established for a monolayer graphene with a staggered sublattice potential. Ref. \cite{Xiao2007} derived the valley-dependent magnetization in a gapped Dirac band, arriving at a relation between the orbital magnetization (per spin) and the band Berry curvature $\Omega(k)$.
			\be 
			M = \frac{e}{\hbar} \int \frac{d^2k}{(2\pi)^2}\mu \Omega(k)f(\epsilon_k)
			\ee
		where $\Omega(k)$ is of opposite signs in the particle and hole bands, $f(\epsilon_k)$ is the Fermi-Dirac distribution. However, as argued in Ref.\cite{Xiao2007}, this relation between $M$ and $\Omega(k)$ holds for a generic band with particle-hole symmetry. 
		
		Applying this formalism to quadratic Dirac band model yields the orbital magnetization in Eq.\eqref{eq:M}.	
	We note that the dependence in Eq.\eqref{eq:M}, with $M$ being constant in each band, is a unique property of the quadratic Dirac band dispersion, Eq.\eqref{eq:H_BBG}. 
	%\cite{SI}. 
	A more general band dispersion would yield $M$ that depends on doping in each band.  These points are further discussed in Appendix \ref{Appendix}, where, for illustration, the orbital magnetization given in Eq. \eqref{eq:M} is derived without invoking the Berry phase. 
% 	for derivation and discussion)
	
	For a fully spin-polarized state at a carrier density $n$, $M_+ = \frac{eD}{2\pi \hbar c}$ whereas $M_-$ depends on density. To calculate $M_-$ we first calculate the chemical potential using $\mu+h = \sqrt{D^2+ (\frac{4\pi n}{2m})^2}$, which gives: 
	\be
	\mu = \sqrt{D^2+ (2\pi n/m)^2} - U(0) \frac{n}2 \sim D - U(0)\frac{n}2 
	,
	\ee
	where the terms $O(n^2)$ were ignored since we are interested in the low-density regime. Plugging this into Eq.\eqref{eq:M} yields:
	\be
	M_- = M(\mu-h) =  
	\lp D - U(0)n\rp \frac{e}{2\pi \hbar c}
	.
	\ee
	As a result, the quantity $\Delta M=M_+-M_-$  equals
	\be
	\Delta M = U(0)n\frac{e}{2\pi \hbar c}
	.
	\ee
	%\addQ{[This expression does not depend on $c$. Is this a typo?]}
	Plugging this into Eq.\eqref{eq:threshold} and using an estimate for spin stiffness obtained above,
	%$J=\frac{\xi^2 h^2}{U(0)}$
    $ J \sim \frac{\xi^2 h^2}{U(0)}$, we find a condition for the transition from a fully polarized state to a chiral spin state:
	\be
	\frac{eU(0)n}{2\pi \hbar c}\geq \frac{2e}{\hbar c} %\lp
	 \frac{\xi^2U(0)n^2}{4}% + \frac{a}{U^2(0)}\rp
	 .
	\ee
This condition can be expressed in terms of carrier density and the correlation length $\xi$ introduced in Sec. \ref{sec:model}:
%	which gives
	\be\label{eq:threshold2}
	n\xi^2 %+ \frac{8a}{nU^3(0)}
	\leq \frac{1}{\pi} 
	.
	\ee
	%\addQ{As a reminder, we have assumed that the $a$ term in $J$ is subleading, so below we focus on the first term on the left-hand side.} 
	Since our mean-field analysis works for $\xi$ exceeding the Fermi wavelength $\lambda_F$,
%	\addQ{If assuming the characteristic lengthscale $\xi$, which is defined through Eq.\eqref{eq:U(q)}, is comparable to Fermi wavelength $1/k_F$, then} 
	the condition in Eq.\eqref{eq:threshold2} is marginally met.
	%\addQ{However,} 
	The threshold Eq.\eqref{eq:threshold2} can be further softened in multilayer graphene (such as tilayer, quadlayer, or pentalayer) since $\Delta M$ is proportional to the valley Chern number. For $N$-layer graphene, in a simplest model, $C$ can take values that scale with the number of layers, 
	$C=N/2$. As a result, the threshold softens to 
	\be\label{eq:criterion_N-layer}
	n\xi^2\leq \frac{C}{\pi}
	.
	\ee
	Using these results, we can predict schematic phase diagram in which the chiral phase (skyrmions) coexists with a uniformly spin-polarized phase, as shown in Fig.\ref{fig:1}.

	%To show that the chiral phase is readily accessible %in realistic systems, 
	%\sout{\addQ{
		%To compare the energies of the chiral and  uniformly spin-polarized phases,
	%To obtain the condition for parameters in realistic systems to achieve chirality,} 
	%we estimate the required size of $\xi$ using realistic parameters.}
Applying the condition for instability toward skyrmion texture to realistic systems requires estimating the values of $\xi$ obtained from microscopic parameters.
	In system of interest, the carrier density $n_c$ at the onset of Stoner transition can be estimated from Eq.\eqref{eq:Stoner transition}. For interaction strength $U(0)=5\times 10^3$meV nm$^2$, a band mass $m=0.03m_e$ 
	and a typical high value 
	\cite{Zhou2022,Seiler2021,de la Barrera2022,Zhou2021isospin,Zhou2021SC} $D=100$meV, 
	it predicts Stoner instability at $n_c \sim 3\times10^{11}$ cm$^{-2}$. This carrier density indeed lies within the density range where the Stoner instability is observed \cite{Zhou2022,Seiler2021,de la Barrera2022,Zhou2021isospin,Zhou2021SC}.
	Eq.\eqref{eq:threshold2} then predicts that to nucleate skyrmions %\sout{the interaction radius}
%\sout{	\addQ{the characteristic length $\xi$} must satisfy $\xi\lesssim 15 \rm{nm}$, a realistic value comparable to \addQ{the screening length of Coulomb interaction in realistic settings. }% the Fermi wavelength.  
%\addQ{ To further stabilize the chiral spin texture, one can to reduce the spin stiffness $J$, which can be achieved by suppressing the correlation length of the spin-spin exchange interaction.}}
		Eq.\eqref{eq:threshold2} then predicts that to nucleate skyrmions the characteristic length $\xi$ must satisfy $\xi\lesssim 15 \rm{nm}$, a value comparable to the screening length of Coulomb interaction in realistic settings. To further stabilize the chiral spin texture, one can reduce the spin stiffness $J$, which can be achieved by suppressing the correlation length of the spin-spin exchange interaction.
	
	A phase diagram describing the competition between the orders described above is shown in Fig.\ref{fig:1}. The transition line from uniformly polarized phase to skyrmion phase is given by Eq.\eqref{eq:threshold2}. We note that, compared to the transition line between the uniformly polarized phase and unpolarized phase, the transition line between skyrmion phase and unpolarized phase is pushed slightly into the unpolarized phase. This is because when skyrmion condenses, the energy contribution from pseudo magnetic field Eq.\eqref{eq:Fc} is always negative and tends to stabilize ordered state. This phase boundary is a first-order phase transition because the translation symmetry and the spin $SU(2)$ symmetry are simultaneously broken on this line.
	
	We note that the condition for skyrmion instability Eq.\eqref{eq:threshold2} can be softened in Dirac bands with larger valley Chern numbers, since $\Delta M$ is proportional to the total Hall conductivity in the lower band. Large valley Chern numbers can be achieved in graphene multilayers, such as trilayer, quadlayer, or pentalayer. 
	Another appealing system is moir\'e graphene, where valley-Chern minibands  \cite{Song2015} 
	give rise to a doping-dependent orbital magnetization, potentially leading to a skyrmion instability triggered by spin polarization onset.
	
	The emergence of skyrmions through the mechanism discussed above can lead to different ground states depending on the interactions between skyrmions and the strength of the spin order parameter zero-point or thermal fluctuations. Strong repulsive interactions would stabilize a chiral skyrmion crystal state, whereas strong fluctuations would lead to a chiral skyrmion liquid. %, arising when zero-point or thermal fluctuations are weak and strong, respectively. 
Overall, these phases are expected to have properties similar to those of the vortex lattice and vortex liquid phases in superconductors \cite{Huberman1979, Fisher1991,Hu1993,Blatter1994}.
	Which of the two states---skyrmion crystal or skyrmion liquid---wins in the true ground state 
	is an interesting topic for future work.
	
	These two phases 
	can be readily distinguished by transport measurements. In the presence of a valley polarization, 
	%\sout{which is ubiquitous} 
	such as the one seen in Bernal bilayer graphene as well as rhombohedral multilayer graphene systems and moire graphene systems, we expect 
	quantized topological Hall effect (QHE) in both states in the absence of an applied magnetic field 
	\cite{Hamamoto2015,Gobel2017}. This QHE contribution will appear in addition to (non-quantized) valley Hall effect. 
	
	So far we discussed skyrmions in one valley. If, however, the system is valley-unpolarized but each valley exhibits Stoner spin-polarized order, skyrmions can occur in both valleys. The valleys $K$ and $K'$ are related by time reversal symmetry, 
	which requires opposite signs of the band Berry curvature and associated orbital magnetization, $M_{K}=-M_{K'}$. 
	This imposes a peculiar relation between skyrmion chiralities in the two valleys, which can be understood from the chiral interaction in Eq.\eqref{eq:deltaF_MB},	
		\be\label{eq:deltaF_MB_K+K'}
	\delta F=\int d^2x \lb -\Delta M_{K} B_{K} -\Delta M_{K'} B_{K'}\rb 
	,
	\ee
where $\Delta M_{K}=M_{K,+}-M_{K,-}$ and $\Delta M_{K'}=M_{K',+}-M_{K',-}$, with the plus and minus indicating contributions of majority-spin and minority-spin carriers. Ignoring, at first, the intervalley exchange interaction, the system can be viewed as two identical Stoner problems with Berry curvatures of opposite signs in the two valleys.
In this case, spin polarization directions in the two valleys are completely decoupled. Then, time-reversal symmetry of the Hamiltonian predicts equal degrees of spin polarization in valleys $K$ and $K'$, and 
requires $M_{K,+} = -M_{K',+}$ and $M_{K,-} = -M_{K',-}$, namely, $\Delta M_+=-\Delta M_-$. Indeed, under time reversal the majority (minority) spins in valley $K$ are always mapped to the majority (minority) spins in valley $K'$. As a result, the ${\rm SU(2)}_{K}\otimes {\rm SU(2)}_{K'}$ symmetry of the microscopic Hamiltonian in the absence of intervalley exchange interactions, enforces the property $\Delta M_+=-\Delta M_-$ regardless of whether spin polarization directions in the two valleys are parallel, antiparallel or canted relative to one another. Therefore, in this case, the system will favor spin textures of opposite chiralities in the two valleys.

In realistic systems, however, the spins in the two valleys are weakly coupled by intervalley exchange interactions, which are much weaker than the intravalley interactions. The intervalley exchange %coupling %should 
	is expected to be of a ferromagnetic sign in the regime of low carrier density \cite{You2022}. To optimize this intervalley interaction, spin textures in valleys $K$ and $K'$ would need to be of identical direction and sign, i.e. $\vec s_K(\vec x)=\vec s_{K'}(\vec  r)$. However, this would result in equal-sign 
	chirality densities in valleys $K$ and $K'$. Such spin textures do not optimize the chiral interaction in Eq.\eqref{eq:deltaF_MB_K+K'} which favors opposite chiral densities in valleys $K$ and $K'$. Therefore,
	%\addQ{On the other hand, the most natural chiral magnetic order favored by chiral interaction features an opposite spin polarization in two valleys $S_K(r) = -S_{K'}(r)$ so that the chirality in valleys $K$ and $K'$ are of opposite sign. Therefore, intervalley interaction disfavors such order.} \sout{and therefore {\bf disfavor the chiral magnetic order with opposite chirality in valleys $K$ and $K’$.}} 
%In that case, 
in this case, our system is expected to exhibit a frustration effect --- the energies of all terms in the Hamiltonian cannot be simultaneously optimized.	%\addLL{[What is the picture of this opposite-chirality phase? How does equal-sign spin polarization in K and K' contradicts opposite chirality? ]} 
As a result, many states can be envisaged as candidates for the ground state. One possibility is a two-valley chiral liquid where spin long-range order is washed out, but the chirality densities being of opposite signs in two valleys and exhibiting a long-range order.   
%\addLL{[Do you mean chirality being  of opposite signs in the two valleys?]} \addQ{[Yes.]}
%\addLL{[Can this phase be called ``valley AFM'' induced by chiral interaction and skyrmions?]} \addQ{OK. Maybe ``non-coplanar valley AFM" to distinguish from collinear ones?}
Another interesting option is a staggered skyrmion lattice, in which skyrmions formed by electrons in valleys $K$ and $K'$ are arranged in two interpenetrating lattices. 
%\sout{ that are shifted in real space relative to each other, instead of sitting right on top of each other.}
 Understanding such frastrated skyrmion phases represents an interesting direction for future work. 

	When time-reversal is not spontaneously broken (no valley polarization), 
	one expects a quantized topological valley Hall effect but no charge Hall effect, %\addQ{[Is this a typo?]} \addQ{[I think the current text is right.]} 
	since the time-reversal symmetry requires skyrmions in valleys $K$ and $K'$ to have opposite chiralities. 
	%\sout{On the other hand,} 
	In addition, the longitudinal transport will be very different in the two phases---vanishing for skyrmion crystal and nonzero for skyrmion liquid, dual to that of superconducting vortex crystals and liquids. 
	
	\section{some implications of skyrmion textures} %conclusions}
	
	In conclusion, this work predicts a geometric spin-orbit coupling that arises in
spin-polarized bands endowed with Berry curvature.  The mechanism underpinning this coupling is that a spin of an electron moving through a spin texture is rotated in spin space. This spin rotation effect, arising due to an electron spin being locked to the local spin quantization axis and tracking it along the electron trajectory, is described by a spin-dependent geometric phase. The adiabatic regime in which the geometric phase picture applies occurs when the Stoner spin gap is large compared to $\hbar v/\ell$, where $\ell$ is the characteristic spatial lengthscale of the spin texture modulation and $v$ is Fermi velocity.

This coupling leads to an instability of a uniformly spin-polarized state towards skyrmions. Occurring in an itinerant magnetic system, the skyrmions have several interesting properties. One is that they act on electrons as a geometric pseudomagnetic field, such that each skyrmion effectively generates one flux quantum of the field. The effective strength of this field is proportional to skyrmion density $n_{\rm s}$ and can be expressed as 
\be\label{eq:Beff}
B_{\rm eff}= 4.13\cdot 10^{-11} \times n_{\rm s} [{\rm cm^{-2}}] \,{\rm Tesla}
.
\ee
For skyrmion density of $n_{\rm s} \approx 10^{10}\,{\rm cm^{-2}}$ this predicts $B_{\rm eff}$ on the order $0.5$ Tesla. The field $B_{\rm eff}$  grows rapidly as $n_{\rm s}$ increases. The Landau levels induced by $B_{\rm eff}$ introduce a new energy scale which governs the skyrmion-induced topological gap in the system spectrum.

The geometric field  $B_{\rm eff}$ would result in a topological Hall effect, manifested through a nonvanishing Hall conductivity occurring in the absence of an applied magnetic field. 
%Since the geometric magnetic field is of opposite sign for carriers with opposite spins, this Hall conductivity will have a characteristic dependence on spin polarization, which will make it easy enough to distinguish from the usual charge Hall conductivity. 
Because the geometric magnetic field has opposite signs for carriers with opposite spins, this Hall conductivity will exhibit a characteristic dependence on spin polarization, distinguishing it from the typical charge Hall conductivity.
%At carrier densities corresponding to $\nu$ electrons per skyrmion, where $\nu$ is an integer, the system will host $\nu$ fully filled skyrmion-induced Landau levels and feature a quantized Hall conductivity of 
At carrier densities corresponding to $\nu$ electrons per skyrmion, where $\nu$ is an integer, the system will host $\nu$ fully filled skyrmion-induced Landau levels, leading to a quantized Hall conductivity
\be
\sigma_{\rm H}=\nu \frac{e^2}{h}.
\ee  
%[{\bf{Say a bit more? Discuss phase diagram?}}] 
For skyrmion crystal or liquid of a high density such that the number of electrons per skyrmion is small, this scenario predicts a state with large $B_{\rm eff}$ and a large topological gap. 

%, featuring an anomalous quantized Hall conductivity. 
%In the extreme limit, when the skyrmion density achieved through this mechanism is large enough to be close to the density of itinerant carriers, the system can spontaneously choose an insulating  ground state where all electrons populate one or several lowest Landau levels. 
%This scenario may be potentially applicable to the quantized Hall phases recently observed in pentalayer graphene in the absence of magnetic field\cite{Han2024,Lu2024}.
%The predicted phase diagram (Fig.\ref{fig:1}), in which the chiral spin texture emerges at low carrier density, aligns with the density regime where quantized Hall effect is observed in pentalayer graphene.
%The predicted dependence on the number of layers, which implies that more layers tend to lower the threshold for the emergence of spin chirality, potentially explains why this effect occurs in pentalayer rather than in bilayer and trilayer.

In the extreme limit, when the skyrmion density achieved through this mechanism is large enough to be close to the density of itinerant carriers, the system can spontaneously adopt an insulating ground state where all electrons occupy one or several lowest Landau levels. This scenario may potentially apply to the quantized Hall phases recently observed in pentalayer graphene in the absence of a magnetic field\cite{Han2024,Lu2024}. The predicted phase diagram (Fig.\ref{fig:1}), in which the chiral spin texture emerges at low carrier density, aligns with the density regime where the quantized Hall effect is observed in pentalayer graphene. The predicted dependence on the number of layers, implying that more layers tend to lower the threshold for the emergence of spin chirality, potentially explains why this effect occurs in pentalayer rather than bilayer and trilayer systems.

%This
%interaction realizes the chiral Heisenberg model which enjoys
%unique advantages such as scale invariance, SU(2) symmetry,
%and hosting an analytically solvable skyrmion ground
%state. We show that this interaction leads to proliferation of
%skyrmions—topologically-protected solitons—without resorting
%to spin-orbital coupling or an applied B field, and therefore
%generates anomalous quantum oscillations. Importantly, unlike
%interactions in ordinary chiral magnets, this chiral interaction
%does not rely on microscopic spin-orbit couplings. Therefore,
%the scenario described in our work is generic and readily applicable
%to all metallic magnets with Berry curvature, including
%moiré graphene and graphene multilayers, which are the focus
%of current experimental efforts.
	
	This work greatly benefited from discussions with Eli Zeldov, Steven Kivelson and Patrick Lee.
	We acknowledge support from the Science and Technology Center for Integrated Quantum Materials, National Science Foundation Grant No. DMR1231319. ZD is currently affiliated with California Institute of Technology, Physics Department.

	\appendix
	\section{Valley-dependent orbital magnetization in graphene bilayer}
	\label{Appendix}

To gain more insight into the physics of the orbital magnetization, Eq.\eqref{eq:M}, 
here we rederive this known result\cite{Xiao2007} by a method that does not explicitly use Berry curvature. Our plan is to calculate the orbital magnetization in an individual graphene valley using thermodynamic relation:
\be\label{eq:def_M}
M_K = -\frac{\partial \Xi_K}{\partial B},
\ee
where $\Xi_K$ is the thermodynamic potential of electrons in this valley, defined as
\be\label{eq:Xi_alpha}
\Xi_K = \sum_\alpha (\epsilon_\alpha-\mu) f(\epsilon_\alpha)
,\quad
f(\epsilon)=\frac1{e^{\beta(\epsilon-\mu)}+1}
\ee
where $\epsilon_\alpha$ are the Landau level energies in the particle and hole bands, labeled by $\alpha=\{\pm,n\}$.

In order to obtain the magnetization at $B=0$ we first calculate the Landau level energies $\epsilon_\alpha$ and, by using the Euler-Maclaurin summation formula, extract the part of the sum over $\alpha$ in Eq.\eqref{eq:Xi_alpha} which is linear in $B$ at small $B$. As we will see, the contribution linear in $B$ is equal to that originating from the anomalous Landau levels reduced by a factor of two, as discussed below. We will end this section by discussing the general character of this result and its relation to the spectral flow. 

The Landau level energies can be derived directly from the BBG Hamiltonian\cite{McCann2013,McCann2006}. For illustration, here we do it for a simplified form of the Hamiltonian involving no trigonal warping terms: 
\be\label{eq:Sup_full_HK}
H _{K}(p)=\lp
\begin{array}{cc}
	D+ \frac{p^2}{2m_0} +\frac{p^2}{2m_a} & -\frac{(p_1- i p_2)^2}{2m}\\
	-\frac{(p_1+ i p_2)^2}{2m} & -D -\frac{p^2}{2m_0} +\frac{p^2}{2m_a}
\end{array}\rp
\ee
Magnetic field can be incorporated in the Hamiltonian through the substitution $\vec p\to\vec p-\frac{e}{c}\vec a$. 
We will first carry out the analysis ignoring the terms $p^2/2m_0$ and $p^2/2m_a$. 
This is justified because these two terms are subleading for a realistic BBG band\cite{McCann2013}. For the same reason we ignore the trigonal warping term (not shown in Eq.\eqref{eq:Sup_full_HK}).
To illustrate the generality of our results, we will subsequently present the analysis for the full Hamiltonian in Eq.\eqref{eq:Sup_full_HK}, finding that the quadratic terms $p^2/2m_0$ and $p^2/2m_a$ do not affect the result. 

Next we consider the Landau levels formed in the presence of a $B$ field, at first excluding the quadratic terms in the diagonal elements. 
As is well known, in each valley --- $K$ or $K'$ --- the Hamiltonian in Eq.\eqref{eq:Sup_full_HK}, with the quadratic terms excluded, in the presence of a magnetic field generates three groups of Landau levels: (i) a pair of anomalous Landau levels at the edges of the hole band for valley $K$ and particle band for valley $K'$, and (ii) two sequences of Landau levels in the particle and hole bands that are related by particle-hole symmetry. 
The energies of these Landau levels in valley $K$ 
can be written as 
\cite{Sup_Koshino2010}
\begin{align}
&\epsilon_{\pm,n} = \epsilon_\pm(x_n)= \pm  \sqrt{x_n^2-\frac{1}{4}\hbar^2 \omega_c^2 +D^2}, 
\quad n\geq2,\nonumber
\\ \label{eq:x_n}
&\epsilon_{0,1}= -D, 
\quad x_n= \hbar \omega_c \lp n-\frac{1}{2}\rp
,
\end{align}
where $\omega_c = eB/mc $ is the cyclotron frequency. For valley $K'$ similar expressions arise, however the anomalous Landau levels are positioned at the particle band edge, $\epsilon_{0,1}= D$. 

Accordingly, the thermodynamic potential $\Xi_K$ in the presence of a $B$ field is 
a sum of three contributions 
\be\label{eq:Xi}
\Xi_K = \Xi_+ +\Xi_- +\Xi_{01},
\ee
where 
\begin{align}
&\Xi_{\pm} = \frac{eB}{h  c}\sum_{n} (\epsilon_{\pm, n}-\mu)f(\epsilon_{\pm, n}),\label{eq:Xi_pm} \\
&\Xi_{01} = \frac{2eB}{hc}(-D-\mu) f(-D)
,
\label{eq:Xi_0}
\end{align}
with $eB/hc$ representing the numbers of electrons in each Landau level per unit area.

Magnetization is given by the linear ($O(B)$) term in $\Xi_K (B)$.
The $O(B)$ contribution from the anomalous levels in each valley is already clearly written in Eq.\eqref{eq:Xi_0}. To calculate the $O(B)$ contribution from $\Xi_\pm$ we use the Euler-Maclaurin formula which approximates a sum by an integral. 
For the contribution of the particle band we have 
\begin{align}\label{eq:Euler_MacLaurin}
&\Xi_+ = \frac{eB}{hc} \lb \frac{1}{\hbar \omega_c} \int_{x_{n=2}}^\infty dx (\epsilon(x)-\mu)f(\epsilon(x)) \right.
\\
&\left.+ \frac{1}{2} (\epsilon\lp x_{n=2}\rp -\mu)f \lp \epsilon(x_{n=2})\rp \rb +O(B^2), \nonumber
\end{align}
where $x_{n=2} = \frac{3}{2}\hbar\omega_c$, see Eq.\eqref{eq:x_n}. Here we have used $\epsilon(\infty)f(\infty)=0$. Working out the integral gives
\be
\Xi_+ 
=-\frac{eB}{hc}(D-\mu)f(D) + O(B^2). \label{eq:Xi+}
\ee
Similarly, the $O(B)$ contribution of the lower-band Landau levels is given by
\be
\Xi_- = -\frac{eB}{hc}(-D-\mu)f(-D) +O(B^2). \label{eq:Xi-}
\ee
After plugging Eqs. \eqref{eq:Xi_0}, \eqref{eq:Xi+} and \eqref{eq:Xi-} into Eqs. \eqref{eq:def_M} and \eqref{eq:Xi}, we arrive at 
\be\label{eq:Sup_M}
M_K(\mu) 
= \begin{cases}
	\frac{2eD}{2\pi \hbar c},\quad \mu>D\\
	\frac{e(\mu+D)}{2\pi \hbar c},\quad -D<\mu<D\\
	0,\quad \mu<-D\\
\end{cases}
.
\ee
We note that this %$M_K$  vs. $\mu$ 
dependence differs by a constant shift %by
of $\Delta M_K = \frac{eD}{2\pi \hbar c}$ from the result in Eq.\eqref{eq:M} that was inferred from the general expression for orbital magnetization obtained in Ref.\cite{Xiao2007}.
%identical to that found in Ref.\cite{Sup_Xiao2007}.
This constant shift arises from the way the contribution of the deep-lying levels is cut off, which is different from the conventional way\cite{Xiao2007}. However, this difference is immaterial because the deep-lying states, due to their uncertain valley character and identical occupancies for opposite spins, %spin-up and spin-down polarization, 
are not expected to affect physical observables. %, since physically one expects that this shift does not affect the result. 

Indeed, at the bottom of the graphene band the carrier states cannot be unambiguously identified with the $K$ and $K'$ valleys. Therefore the ambiguity arising from the cutoff is a matter of convention rather than a physical effect.  
%since it only signifies an ambiguity of where to cutoff ``valley K". 
%The irrelevance of this constant shift can also be seen explicitly by noting that 
Furthermore, the quantity that %finally enters our theory 
matters for the physics of interest is the difference of the contributions from the spin-up and spin-down bands, $\Delta M=M_{K,\uparrow}-M_{K,\downarrow}$. The bands for opposite spins are filled equally at the bottom, such that the contributions of the deep-lying states to $M_{K,\uparrow}$ and $M_{K,\downarrow}$ cancel each other.
%which suggest that the ambiguity due to high-energy cutoff enters both $M_{K,\uparrow}$ and $M_{K,\downarrow}$ and cancels each other.
%\sout{The absence of a particle-hole symmetry is due to the fact that the orbital magnetization $M_K(\mu)$ is not required to be particle-hole symmetric, since $M_K$ in an empty valley has to vanish, whereas a fully occupied valley is allowed to have a nonvanishing $M_K$.}
%\be\label{eq:M_final}
%M(\mu) = \begin{cases}
%	\frac{2eD}{2\pi \hbar},\quad \mu>D\\
%	\frac{e(\mu+D)}{2\pi \hbar},\quad -D<\mu<D\\
%	0,\quad \mu<-D\\
%\end{cases}
%\ee

The meaning of the resulting dependence $M_K(\mu)$, %in Eq.\eqref{eq:M_final} 
in which $M_K$ is constant when the Fermi level lies within one of the bands, can be understood in terms of a spectral flow induced by a variation of $B$. Namely, the role of the Landau levels moving up and down is merely to cancel half of the contribution to magnetization $M_K$ of the anomalous Landau levels in the corresponding bands. As a result, there is no $\mu$ dependence when the Fermi level %is within each band, 
lies outside the gap. In that each anomalous level contributes a half of the `nominal value' of a single Landau level. This contribution comes with a plus sign or a minus sign depending on whether an anomalous Landau level is present or absent for the band and valley in question. The resulting dependence of orbital magnetization is identical for the $K$ and $K'$ valleys up to a sign reversal, $M_K(\mu)=-M_{K'}(\mu)$. 

This analysis can be applied to a realistic model of Bernal bilayer graphene, where the band Hamiltonian takes a more complicated form\cite{McCann2013}. Here we show that adding the two quadratic terms given in Eq.\eqref{eq:Sup_full_HK}, that were neglected temporarily, does not alter the result for $M_K(\mu)$.

The term $p^2/2m_a$ is an identity matrix in the sublattice variables. As a result, it merely shifts the energy eigenvalues without affecting the electron wavefunction that determines the orbital magnetization. Therefore, this term only affect the diamagnetic susceptibility but does not affect the magnetization at $B=0$. Indeed, adding it in Eq.\eqref{eq:Euler_MacLaurin} yields an $O(B^2)$ contribution to the themodynamic potential, changing somewhat the diamagnetic susceptibility but not changing $M_K(\mu)$. 

%, we find that adding it in Eq.\eqref{eq:Euler_MacLaurin} only yields an $O(B^2)$ contribution to the themodynamic potential. Therefore, this term only affect the diamagnetic susceptibility but does not affect the magnetization at $B=0$. A physical reason is that the term $p^2/2m_a$ is an identity matrix, which merely shifts the energy without affecting the electron wavefunction that determines the orbital magnetization. 
%As a result, it does not affect the rotation behavior of electrons. %, and therefore does not affect the magnetization. 

The term $p^2/2m_0$ has a $\sigma_3$ sublattice structure. Such a term does affect the wavefunctions, and yet, this term alone does not break the particle-hole symmetry. Also, this term does not affect the energy of the lowest Landau level in the particle band. As a result, the two conditions necessary for the reasoning above [from Eq.\eqref{eq:Xi} to Eq.\eqref{eq:Xi-}] --- the particle-hole symmetry and the presence of two anomalous Landau levels --- remain valid. As a result, the answer for magnetization given above remains unchanged.

\end{document}